\documentstyle[prl,aps]{revtex}
\large
\begin{document}
\twocolumn

\draft

\title{The Influence of Superpositional Wave Function Oscillations \\
on Shor's  Quantum Algorithm}

\author{{ Gennady P. Berman$^1$, Gary D. Doolen$^1$, and Vladimir I. Tsifrinovich$^2$}}

\address{
$^1$Theoretical Division and CNLS, Los Alamos National Laboratory, 
Los Alamos, New Mexico 87545\\
$^2$Department of Applied Mathematics and Physics, Polytechnic University,
Six Metrotech Center, Brooklyn NY 11201}

\maketitle

\begin{abstract} 
We investigate the influence of superpositional wave function oscillations on the performance of Shor's quantum algorithm for factorization of integers. It is shown that the wave function oscillations can destroy the required quantum interference. This undesirable effect can be routinely eliminated using a resonant pulse implementation of quantum computation, but requires special analysis for non-resonant implementations.       
\newline
\renewcommand{\baselinestretch}{1.656}

%
\end{abstract}
\quad\\
{\bf I. Introduction}\\ \ \\
The recent intensive interest in quantum computation was stimulated by Shor's discovery of a quantum algorithm for the factorization of integers \cite{shor1}. Recent theoretical quantum computation achievements include: universal quantum logic gates \cite{bare}, quantum error correction codes \cite{shor2,steane}, Grover's algorithm for fast quantum searches \cite{grov}, and implementation of quantum algorithms using a statistical ensemble of nuclear spins at room temperature \cite{chuang1}-\cite{cory}. Introductory discussions on quantum computation are presented in \cite{will,ber}. One problem which has not yet been discussed is the influence of wave function oscillations on the performance of quantum algorithms. To carry out a quantum computation, one utilizes superpositions of eigenstates of the Hamiltonian. As a result, each state can acquire its own phase, causing destructive and constructive interference. This interference is essential for quantum computation.

In this paper we demonstrate the effect of wave function oscillations on quantum interferences, using a simple example of Shor's quantum algorithm for factoring the number 4. We show that fast oscillations can destroy required interference effects. We explain how this effect can be routinely eliminated for resonant pulse implementation of quantum computation. We also point out that,  for ``non-resonant'' proposals for quantum computation, the effects of wave function oscillations will require additional consideration.
\\ \ \\
{\bf II. A Model System of Four Qubits}\\ \ \\
Consider four qubits which are prepared initially in the ground state. We shall use these qubits to implement  Shor's quantum algorithm. The two left-most qubits represent the number ``x'', and  the two right-most qubits represent ``y''. Assume that initially all four qubits are in their ground states. The wave function of the system is,
$$
\Psi(0^-)=|0_10_0,0_10_0\rangle,\eqno(1)
$$
where $0^-$ denotes a moment of time, $t=0-\epsilon$, where $\epsilon\rightarrow 0$; each subscript denotes the position of a qubit in a register; $|0\rangle$ and $|1\rangle$ denote, correspondingly, the ground state and the excited state of a qubit. 

To demonstrate how wave function oscillations can destroy the desired interference in Shor's algorithm, we use in this section a simplified consideration. Namely, we assume that Shor's algorithm can be implemented using three unitary transformations: The first transformation produces a superposition of all possible states of the $x$-register; the second transformation ``computes'' the periodic function, $y(x)$; and the third transformation performs a Discrete Fourier Transform (DFT). Next, we assume that all three transformations have infinitesimally short time duration, but there exist finite time intervals between transformations. A more realistic model will be discussed in section III.

Let us now factor a smallest composite number, $4$, using a four-qubit quantum computer. (Generally, to factor a number $N$, one should have the number of states in the $x$-register greater than $N^2$ but it is not important for our consideration.)  Following Shor's method, we choose the only coprime number, $3$. The quantum computer then must compute the periodic function,
$$
y(x)=3^x~(mod~4).\eqno(2)
$$
This function is,
$$
y(0)=1,\quad y(1)=3,\quad y(2)=1,\quad y(3)=3,...
$$
Then, the quantum computer must find the period of this function. ($T=2$, in this case.) The factor of 4 can be found as the greatest common divisor (GCD) of two numbers: $(z+1)$ and 4 or $(z-1)$ and 4, where $z=3^{T/2}$. In our case, 
$$
GCD(z-1,4)=GCD(2,4)=2,\eqno(3)
$$
gives the desired factor of 4. 
We shall use decimal notation below for the states of the registers,
$$
|m_1m_0,n_1n_0\rangle\rightarrow |m,n\rangle,~m=m_0+2m_1,~n=n_0+2n_1.\eqno(4)
$$
In this notation the initial wave function (1) has the form: $|0,0\rangle$. 

Now let us consider the sequence of three quantum transformations required to compute the period, $T$. The first quantum transformation creates a superposition of all possible values of $x$, with a $y$-value of zero,
$$
\Psi_1={{1}\over{2}}(|0,0\rangle+|1,0\rangle+|2,0\rangle+|3,0\rangle).
\eqno(5)
$$
Assume that the delay time between the first and the second transformations is $\tau_1$. According to the Schr\"odinger equation, during the time interval $\tau_1$ each state in (5) acquires a phase, $-E_{pq}\tau_1$, where $E_{pq}$ is the energy of the state $|p,q\rangle$ and we put $\hbar=1$. After time $\tau_1$, the wave function takes the form,
$$
\Psi_1(\tau_1^-)={{1}\over{2}}\Bigg(|0,0\rangle e^{-iE_{00}\tau_1}+ |1,0\rangle e^{-iE_{10}\tau_1}+\eqno(6)
$$
$$
|2,0\rangle e^{-iE_{20}\tau_1}+ |3,0\rangle e^{-iE_{30}\tau_1}\Bigg).
$$
The second transformation computes the values of the periodic function (2). Thus, one gets from (6) the wave function,
$$
\Psi_1(\tau^+_1)={{1}\over{2}}\Bigg(|0,1\rangle e^{-iE_{00}\tau_1}+ |1,3\rangle e^{-iE_{10}\tau_1}+\eqno(7)
$$
$$
|2,1\rangle e^{-iE_{20}\tau_1}+ |3,3\rangle e^{-iE_{30}\tau_1}\Bigg),
$$
At this point, the ``natural'' relation between the state and its phase is destroyed: the phase factor of the state $|m,n\rangle$ is not equal to $-E_{mn}\tau_1$.

During the next time interval, $\tau_2$, the wave function evolves further according to the Schr\"odinger equation and takes the form,
$$
\Psi(\tau_1+\tau_2^-)=
{{1}\over{2}}\{|0,1\rangle\exp(-iE_{00}\tau_1-iE_{01}\tau_2)+\eqno(8)
$$
$$
|1,3\rangle\exp(-iE_{10}\tau_1-iE_{13}\tau_2)+
$$
$$
|2,1\rangle\exp(-iE_{20}\tau_1-iE_{21}\tau_2)+
|3,3\rangle\exp(-iE_{30}\tau_1-iE_{33}\tau_2)\}.
$$
Finally, at $t=\tau_1+\tau_2$ one performs the DFT of the $x$-register,%
$$
|x\rangle\rightarrow{{1}\over{2}}\sum_{k=0}^3e^{2\pi ikx/4}|k\rangle.\eqno(9)
$$
Then, the wave function takes the form,
$$
\Psi(\tau_1+\tau_2^+)={{1}\over{4}}[(|0,1\rangle+|1,1\rangle+|2,1\rangle+\eqno(10)
$$
$$
|3,1\rangle)
\exp(-iE_{00}\tau_1-iE_{01}\tau_2)+ 
$$
$$
(|0,3\rangle+i|1,3\rangle-|2,3\rangle-i|3,3\rangle)
\exp(-iE_{10}\tau_1-iE_{13}\tau_2)+
$$
$$
(|0,1\rangle-|1,1\rangle+|2,1\rangle-|3,1\rangle)
\exp(-iE_{20}\tau_1-iE_{21}\tau_2)+ 
$$
$$
(|0,3\rangle-i|1,3\rangle-|2,3\rangle+i|3,3\rangle)
\exp(-iE_{30}\tau_1-iE_{33}\tau_2)].
$$
One can see the destructive impact of the wave function oscillations on the quantum computation. (The phase advance at later times $t>\tau_1+\tau^+_2$ does not further change the quantum interference.)  If $\tau_1=\tau_2=0$, one gets the desired constructive and destructive interferences. In this case, the wave function (10) reduces to the following,
$$
\Psi={{1}\over{2}}(|0,1\rangle+|2,1\rangle+|0,3\rangle-|2,3\rangle).\eqno(11)
$$
Measuring the state of the $x$-register one gets two values for $x$: $x_1=0$ and $x_2=2$. The period, $T$, can be found as the ratio, $D/x_2$, where $D$ is the number of states in $x$-register \cite{ekert}. In our case, $D=4$ and $T=4/2=2$. 

If $\tau_1, \tau_2\not=0$, the required quantum interference is destroyed unless the following very special conditions are satisfied,
$$
(E_{20}-E_{00})\tau_1+(E_{21}-E_{01})\tau_2=2\pi n_1,\eqno(12)
$$
$$
(E_{30}-E_{10})\tau_1+(E_{33}-E_{13})\tau_2=2\pi n_2,
$$
where $n_1$ and $n_2$ are arbitrary integers. For example, the state $|1,1\rangle$, which vanishes if $\tau_1=\tau_2=0$, now survives as a sum of two terms,
$$
{{1}\over{4}}|1,1\rangle[\exp(-iE_{00}\tau_1-iE_{01}\tau_2)-\exp(-iE_{20}\tau_1-iE_{21}\tau_2)].\eqno(13)
$$
The first term  $(1/2)|0,1\rangle$ in the superposition (11) is described now by two terms,
$$
{{1}\over{4}}|0,1\rangle[\exp(-iE_{00}\tau_1-iE_{01}\tau_2)+\exp(-iE_{20}\tau_1-iE_{21}\tau_2)].\eqno(14)
$$

The reason Shor's algorithm fails in this example is that the ``like'' terms corresponding to the same state, carry their ``history'' after their initialization. For example, the first term in (14) is generated by the transformation sequence,
$$
|0,0\rangle\rightarrow|0,0\rangle\stackrel{\tau_1}{\rightarrow}|0,1\rangle
\stackrel{\tau_2}{\rightarrow}|0,1\rangle,\eqno(15)
$$
while the second term in (14) is generated by  the transformation sequence,
$$
|0,0\rangle\rightarrow|2,0\rangle\stackrel{\tau_1}{\rightarrow}|2,1\rangle
\stackrel{\tau_2}{\rightarrow}|0,1\rangle.\eqno(16)
$$
One can see that the two different phase factors in (14) directly reflect the ``history'' of the corresponding terms (15) and (16).

To provide a quantum algorithm for  a finite time one must create the same phase factor at the end of the algorithm's implementation, for all like terms.
The natural way to solve this problem is to ``induce'' the ``natural'' phase, $-E_{mn}t$, every time the term $|m,n\rangle$ is generated in the process of quantum computation. In the next section we show that this can be done routinely using a resonant pulse implementation of quantum  computation.\\ \ \\
{\bf III. Generating ``natural'' phase factors using \\resonant pulses}\\ \ \\

Following the Lloyd's idea \cite{l1}, we assume that all quantum transformations are implemented by a proper sequence of resonant pulses. Each pulse drives a quantum transformation between the corresponding energy levels in a system of weakly interacting qubits. Next, we assume that one keeps the stable continuous reference oscillations, $\sim\exp(i\omega_{pk}t)$, with the frequencies, $\omega_{pk}=E_p-E_k$, for each resonant transition which is used in the process of quantum computation. (Here $E_p$ and $E_k$ are the energies of the states $|p\rangle$ and $|k\rangle$ of the whole system of qubits.) The resonant pulses are supposed to be ``cut'' from the reference oscillations and to be applied to the system of qubits with a proper phase shift, $\varphi$ ($-\pi<\varphi\le\pi$) relative to the reference oscillations. This technique is routinely used in nuclear magnetic resonance.

Now we show how a natural phase is generated by a resonant pulse. Let's consider the state $|k\rangle$ of the whole system of qubits. Suppose that the resonant pulse with frequency $\omega_{pk}$ drives the transition from state $|k\rangle$ to state $|p\rangle$, i.e. ``generates'' the state $|p\rangle$. Neglecting non-resonant effects, the Hamiltonian for this process can be written as,
$$
{\cal H}=E_k|k\rangle\langle k|+E_p|p\rangle\langle p|-\eqno(17)
$$
$$
{{\Omega}\over{2}}\Bigg[
e^{i(\omega_{pk} t+\varphi)}|k\rangle\langle p|+
e^{-i(\omega_{pk} t+\varphi)}|p\rangle\langle k|\Bigg].
$$
Here $\Omega$ is the Rabi frequency, and $\varphi$ is the phase of the resonant pulse. 

Assume that at the beginning of the pulse ($t=t_0$) the amplitude of the state $|k\rangle$, $C_k(t_0)$, has the natural phase factor,
$$
C_k(t_0)=|C_k(t_0)|\exp(-iE_kt_0),\eqno(18)
$$
and the amplitude of the state $|p\rangle$, $C_p(t_0)=0$. The Schr\"odinger equation for the amplitudes $C_k$ and $C_p$, for the states $|k\rangle$ and  $|p\rangle$, reduces to,
$$
i\dot C_k=E_kC_k-{{1}\over{2}}\Omega e^{i(\omega_{pk} t+\varphi)}C_p,\eqno(19)
$$
$$
 i\dot C_p=E_pC_p-{{1}\over{2}}\Omega e^{-i(\omega_{pk} t+\varphi)}C_k.
$$
The solutions of these equations are,
$$
C_k(t)=C_k(t_0)\cos[\Omega(t-t_0)/2]\exp[iE_k(t_0-t)],\eqno(20)
$$
$$
C_p(t)=C_k(t_0)\sin[\Omega(t-t_0)/2]\exp[i(\pi/2-\varphi+E_kt_0-E_pt)].
$$
At the end of the pulse: $t=t_0+\tau$, one has from (20),
$$
C_k(t_0+\tau)=|C_k(t_0)|\cos\alpha\exp[-iE_k(t_0+\tau)],~\alpha=\Omega\tau/2,\eqno(21)
$$
$$
C_p(t_0+\tau)=|C_k(t_0)|\sin\alpha\exp[i(\pi/2-\varphi)]\exp[-iE_p(t_0+\tau)].
$$
One can see that at the end of the pulse the newborn state, $|p\rangle$, has the natural phase factor, $\exp(-iE_pt)$, while the initial state, $|k\rangle$, keeps its natural factor, $\exp(-iE_kt)$. This shows that the resonant pulse transfers the phase accumulated by the reference oscillations to the newborn state, $|p\rangle$, providing the natural phase factor, $\exp(-iE_pt)$. The factor $\exp[i(\pi/2-\varphi)]$ in the expression for $C_p(t_0+\tau)$ is the ``standard'' phase shift which can be eliminated by choosing $\varphi=\pi/2$.

On the contrary, let us consider now the case when a resonant pulse is not cut from the reference oscillations, and has the initial phase, $\varphi_0$ $(-\pi<\varphi_0\le\pi)$, at time $t=t_0$. In this case, the interaction between a qubit and a resonant pulse (the third term in (17)) has the form,
$$
-{{\Omega}\over{2}}\{\exp[i\omega_{pk}(t-t_0)+i\varphi_0]|k\rangle\langle  p|+\eqno(22)
$$
$$ 
\exp[-i\omega_{pk}(t-t_0)-i\varphi_0]
|p\rangle\langle  k|\}.
$$
In this case, the amplitude of the newborn state, $|p\rangle$, in (21) will be replaced by,
$$
C_p(t_0+\tau)=|C_k(t_0)|\sin\alpha\exp[i(\pi/2-\varphi_0)]\times\eqno(23)
$$
$$
\exp(-iE_kt_0-iE_p\tau).
$$
Instead of the natural phase (21), one has a phase which reflects the history of the corresponding state. Namely, the factor $\exp(-iE_kt_0-iE_p\tau)$ in (23) shows that the state $|p\rangle$ was generated from the state $|k\rangle$, at $t=t_0$. 
To get the natural phase, one could connect the value of the initial phase, $\varphi_0$, to the time of the pulse application, $t_0$. Indeed, choosing $\varphi_0=\varphi+\omega_{pk}t_0$ (where $\varphi$ is a proper phase shift: $-\pi\le\varphi\le\pi$), we obtain the expression (21) derived for a coherent pulse. However, in the case of application of a non-coherent pulse, the inevitable uncertainty, $\delta t_0$, can cause a large uncontrolled phase shift, $\omega_{pk}\delta t_0$. Taking into consideration that quantum computation requires application of a large number of pulses, one can conclude that practical implementation of non-coherent pulses is a complicated technical problem. 
Thus, applying non-coherent resonant pulses one can destroy the desired quantum interference. 

The same destructive effect may occur for non-resonant implementations of quantum computation. Assume, for example, that a powerful pulse, $V(t)$, 
$$
V(t)=\cases{V, &if $  t_0\le t\le t_0+\tau$;\cr
0,& if $t<t_0$,~$t>t_0+\tau$,\cr}
$$
acts on an individual qubit driving it from the ground state to the excited state (or, vice versa,
  from the excited state to the ground state). This corresponds, for example, to the transition of the whole system of qubits from the state, $|k\rangle$ to the state $|p\rangle$. The last term in (17) which describes the interaction between the qubits and the external field, we present in the form,
$$
-V(t)(|k\rangle\langle p|+|p\rangle\langle k|).\eqno(24)
$$
For the simplest case, $V\rightarrow\infty$,~$\tau\rightarrow 0$,~ $V\tau\rightarrow=\alpha$, we derive the solution,
$$
C_k(t_0^+)=C_k(t_0^-)\cos\alpha,~ C_p(t_0^+)=iC_k(t_0^-)\sin\alpha.\eqno(25)
$$
One can see from Eq. (25) that the newborn state, $|p\rangle$, acquires the phase accumulated by the ``parent'' state, $|k\rangle$ (plus the ``standard'' phase shift, $\pi/2$). Again in this case, the phase of the newborn state, $|p\rangle$ (with the accuracy to the ``standard'' phase shift, $\pi/2$) is equal to $-E_kt_0$, instead of the natural phase, $-E_pt_0$.
\\ \ \\ 
{\bf Conclusions}\\ \ \\
We have demonstrated the importance of wave function oscillations for 
finite-time implementation of Shor's quantum algorithm. These oscillations can destroy the desired interference required for quantum computation. For resonant implementations of quantum computation, the detrimental effects associated  with this phase factor can be routinely eliminated if one can keep stable continuous reference oscillations with the resonant frequency for each quantum transition which is used in the process of quantum computation. For ``non-resonant'' proposals for quantum computation the influence of wave function oscillations on performance of quantum algorithms requires a special consideration.\\ \ \\
{\bf Acknowledgments}\\ \ \\
This work  was supported by the Department of Energy under contract W-7405-ENG-36, and by the National Security Agency.
\newpage

\end{document}